\documentclass[reqno,12pt]{amsart}
\usepackage{amsmath, latexsym, amsfonts, amssymb, amsthm, amscd}
\usepackage{graphicx, color}

\numberwithin{equation}{section}

\newtheorem{theorem}{Theorem}[section]
\newtheorem{lemma}[theorem]{Lemma}
\newtheorem{proposition}[theorem]{Proposition}
\newtheorem{corollary}[theorem]{Corollary}
\newtheorem{rem}[theorem]{Remark}

\renewcommand{\tilde}{\widetilde}          
\DeclareMathSymbol{\leqslant}{\mathalpha}{AMSa}{"36} 
\DeclareMathSymbol{\geqslant}{\mathalpha}{AMSa}{"3E} 
\DeclareMathSymbol{\eset}{\mathalpha}{AMSb}{"3F}     
\renewcommand{\leq}{\;\leqslant\;}                   
\renewcommand{\geq}{\;\geqslant\;}                   

\newcommand{\R}{\mathbb{R}}
\newcommand{\Z}{\mathbb{Z}}

\title[Forecasting volatility with the multifractal random walk model]{Forecasting volatility with the multifractal random walk model}
\author{}

\begin{document}

\maketitle
\begin{center}
{Jean Duchon, Raoul Robert \\
\footnotesize \noindent
 Institut Fourier, universit{\'e} Grenoble 1, UMR CNRS 5582, \\ 
 100, rue des Math{\'e}matiques, BP 74, 38402 Saint-Martin d'H{\`e}res cedex, France}

{\footnotesize \noindent e-mail: \texttt{Jean.Duchon@ujf-grenoble.fr,Raoul.Robert@ujf-grenoble.fr}}

\bigskip

{Vincent Vargas \\
\footnotesize 
 
 CNRS, UMR 7534, F-75016 Paris, France \\
  Universit{\'e} Paris-Dauphine, Ceremade, F-75016 Paris, France} \\

{\footnotesize \noindent e-mail: \texttt{vargas@ceremade.dauphine.fr}}
\end{center}

\begin{abstract}
We study the problem of forecasting volatility for the multifractal random walk model. In order to avoid the ill posed problem of estimating the correlation length $T$ of the model, we introduce a limiting object defined in a quotient space; formally, this object is an infinite range logvolatility. For this object and the non limiting object, we obtain precise prediction formulas and we apply them to the problem of forecasting volatility and pricing options with the MRW model in the absence of a reliable estimate of $\sigma$ and $T$. 

\end{abstract}
\vspace{1cm}
\footnotesize


\noindent{\bf Key words or phrases:} Random measures, Gaussian processes, Prediction theory, Multifractal processes.


\normalsize

\section{Introduction}

In recent years, the Multifractal Random Walk (MRW) model introduced by Bacry, Delour and Muzy (\cite{cf:BaDeMu}) has received much attention from the financial practitioners. The MRW appears as a natural extension of the basic geometric Brownian model (GB).
Let $S_{t}$ be the price of an asset; in the GB model we write:
\begin{equation}\label{eq:GB}
S_{t}=S_{0}e^{Y_{t}},
\end{equation}
where $Y_{t}=(\mu-\frac{1}{2}\sigma^{2})t+\sigma B_{t}$. Here $\mu$ is the mean return rate , $\sigma$ the volatility of the asset supposed to be constant, $B_{t}$ a standard Brownian motion. The GB model plays a fundamental role in finance as it enables to give a price to options (by the famous Black-Scholes formula). Nevertheless when compared with reality the model displays severe drawbacks; more precizely, the GB model does not display the following stylized facts which are largely acknowledged in the litterature (\cite{cf:Cizeau}, \cite{cf:Co}):
 \begin{itemize} 
\item
The volatility fluctuates randomly and follows approximately a lognormal distribution.
\item	
While the returns are rapidly decorrelated, the volatility exhibits long range correlations following a power law
\item
The returns are heavy tailed.
\end{itemize}
The above stylized facts lead naturally to propose as model of volatility a random measure called the limit lognormal model. This object, introduced by Mandelbrot (\cite{cf:Man}) in the context of turbulence, was rigorously defined and studied by Kahane in \cite{cf:Kah} under the name of gaussian multiplicative chaos. More precisely, Kahane developped a general theory for  $\sigma$-positive kernels (see \cite{cf:Kah} for the exact definition) and suggested how to use this framework to define the object introduced by Mandelbrot in \cite{cf:Man}.

Let us give a very sketchy presentation of the gaussian multiplicative chaos. Let $T$ and $\gamma$ be given parameters ($T$ is the correlation length). We consider the positive kernel $\rho(s,t)=\ln^{+}(T/|t-s|)$. To this kernel, we associate the stationary gaussian process $X_{t}^{T}$ with covariance $\rho(s,t)$ (in a generalized sense since $\rho(t,t)=\infty$). The associated multiplicative chaos is a random measure defined (formally) by:
\begin{equation*}     
m(dt)=e^{\gamma X_{t}^{T}-\frac{\gamma^2}{2}E((X_{t}^{T})^2)}dt.
\end{equation*}
Of course, this formula has no meaning since $E((X_{t}^{T})^2)=\infty$ but it clearly suggests the limit procedure by which we can define rigorously the random measure (in \cite{cf:BaMu}, it is proven that the kernel $\rho$ is $\sigma$-positive in the sense of Kahane and therefore the measure $m$ is a particular case of the general theory developped in \cite{cf:Kah}). 

Following an idea of Mandelbrot and Taylor (\cite{cf:ManTay}), one can consider the model (for the log price) of a Brownian motion subordinated by an independent  random mesure. This was first performed with the measure $m$ by the authors of \cite{cf:BaMu}: this defines the MRW model. More precisely, we define the random time change $\theta(t)=m([0,t])$. The MRW model is then the four parameter stochastic process given by:
\begin{equation*} 
Y_{t}^T=\mu t -\frac{1}{2}\sigma^2\theta(t)+\sigma B_{\theta(t)},
\end{equation*}    
where the Brownian motion $B$ is independent of $m$. If $t$ is not too large (a few years for example), the drift term $\mu t -\frac{1}{2}\sigma^2\theta(t)$ is in practice negligible compared to the Brownian term so the MRW reduces to the three parameter  $(\sigma, \gamma,T)$ process:
\begin{equation*} 
Y_{t}^T=\sigma B_{\theta(t)}.
\end{equation*}    
From classical properties of the chaos $m$, it follows that $\theta(t)$ is a continuous process so that $Y_{t}^T$ is also continuous.  
By using the scale invariance of $B$, one can write (at least formally) $Y_{t}^T$ as a stochastic integral:
\begin{equation}\label{eq:defMRW} 
Y_{t}^T=\sigma \int_{0}^{t}e^{\lambda X_{s}^{T}-\lambda^2E((X_{s}^{T})^2)}dB_{s}, 
\end{equation} 
where $\lambda=\frac{\gamma}{2}$ is the intermittency coefficient. One can then obtain $Y_{t}^T$ rigorously as the limit of a discretized approximation scheme; more precisely,  let $\tau$ be some positive step parameter, $(\epsilon_{n})_{n \in \Z}$ be a standard gaussian i.i.d. sequence and  $(X^{\tau}_{n})_{n \in \Z}$ be a zero mean stationary gaussian sequence independant of $\epsilon$ with kernel (see lemma (\ref{lem:app}) of the appendix for the existence of $X^{\tau}$):     
\begin{equation} \label{eq:defiX}
E[X^{\tau}_{n}X^{\tau}_{p}]=\ln^{+}(\frac{T}{(|n-p|+1)\tau}).
\end{equation}   
If $4\lambda^2 < 1$, one can show the convergence in law (in a functional sense) as $\tau$ goes to $0$ of $(Y^{\tau,T}_{t})_{t \geq 0}$ given by:
 \begin{equation*}
Y^{\tau,T}_{t}=\sigma \sqrt{\tau} \sum_{n=1}^{ \lfloor t/ \tau \rfloor }e^{\lambda X^{\tau}_{n}-\lambda^2 \ln^{+}(T/ \tau)}\epsilon_{n}.
\end{equation*}
towards $(Y_{t}^T)_{t \geq 0}$ (\cite{cf:BaMu}).

To what extent is $Y_{t}^T$ relevant to model financial markets? As it gives a fat tailed distribution for the returns, long range correlations and a lognormal fluctuating volatility, the MRW appears as a rather relevant model for exchange markets. The main limitation of this model is the symmetry of the distribution of returns and therefore it can not capture the leverage effect observed on stocks and indices (see \cite{cf:BoMaPo} for a quantitative study of the leverage effect) .
One of the most important predictions of the process  $(Y_{t}^T)_{t \geq 0}$ concerns the logvariogram.
More specifically, let $\tilde{Y}_{j}^T=\underset{t \in [j;j+1]}{\max}Y_{t}^T-\underset{t \in [j;j+1]}{\min}Y_{t}^T$.
Then, one defines the logvariogram $V(j)$ by the formula:
\begin{equation*}
V(j)=E[(\ln(\tilde{Y}_{j}^T)-\ln(\tilde{Y}_{0}^T))^{2}].
\end{equation*}
One can show that there exists $C>0$  such that for all $j \leq T$:
\begin{equation}\label{eq:Logvar}
V(j)=C+2\lambda^2\ln(j)+\lambda^2\delta(j,\lambda^2)
\end{equation} 
where $C \simeq 0.29$ and the terms $\delta(j,\lambda^2)$ do not depend on $T$ or $\sigma$ such that: 
\begin{equation*}
\sup_{j \leq T}\delta(j,\lambda^2) \underset{\lambda^2 \to 0}{\rightarrow}0.
\end{equation*}
In practice, one can neglect the last term of the above expansion since the value of $\lambda^2$ for a financial asset (stock, currency, index, etc...) will typically belong to the interval $[0.01,0.06]$. 

We see on the figures \ref{fig:first} and \ref{fig:second} the empirical logvariogram of the SP500 index on three disjoint periods of 6 years and the empirical logvariogram of the currency Euro/Australian dollar on the period $2001-2007$ along with the corresponding regressions of the form $C+2\lambda^2\ln(j)$. One can notice that the logvariograms of the SP500 are less and less noisy as time evolves (perhaps due to the increase of liquidity: tick size, volume, etc...). We propose to estimate the intermittency parameter $\lambda^2$ by the aformentionned regression: Monte Carlo simulations performed with $1000$ trials of a MRW on a period of $6$ years with $\lambda^2=0.02$ show that the above estimator belongs to the interval $[0.01,0.03]$ with a confidence interval of $90$ percent.      

Finally, we mention the problem of the estimation of the parameter $T$. Typically, $T$ is very hard to estimate precisely (With the financial data available, one finds huge error bars on the estimation of this parameter: for typical estimates of $T$, see \cite{cf:BaKoMu}). Roughly, $T$ will exceed 2-3 years for an asset. It is easy to convince oneself that beyond a window of size $5-6$ years, one can not assume that the sequence of returns of an asset forms a stationary sequence because markets evolve in time. In these conditions, ergodicity of the data breaks down and one can not rely on a precise estimate of $T$. In fact, it is even impossible to determine $T$ with data on a window of size $T$; indeed, if $Y^{T}$ and $Y^{\tilde{T}}$ are two MRW with respective parameters $(\sigma, \lambda,T)$ and $(\sigma, \lambda,\tilde{T})$ such that $T \leq \tilde{T}$ then the following generalized scale invariance holds:
\begin{equation}\label{eq:scaleinv}
(Y^{\tilde{T}}_{t})_{t \in [0,T]} \overset {(law)} {=} e^{\lambda\Omega_{\tilde{T}/T}-\lambda^{2}\ln(\tilde{T}/T)}(Y^{T}_{t})_{t \in [0,T]}, 
\end{equation}
where $\Omega_{\tilde{T}/T}$ is a centered gaussian variable independent from $(Y^{T}_{t})_{t \in [0,T]}$ and of variance $\ln(\tilde{T}/T)$. 
Therefore, the law of $Y^{\tilde{T}}$ given $\Omega_{\tilde{T}/T}=x$ is the law of a MRW with parameters $(\sigma e^{\lambda x-\lambda^{2}\ln(\tilde{T}/T)}, \lambda,T)$.
In conclusion, the statistical estimation of $T$ and $\sigma$ is an ill posed problem and the most natural way to get rid of the problem is thus to let $T$ go to infinity and work with the limiting object. In doing such a procedure, one must work in a quotient space (see definition (\ref{eq:defX}) where one views the log volatility $X$ as an element of the quotient space $\mathcal{S}'(\R)/ \R$).

The rest of the article is organized as follows: in section 2, we give some preliminary definitions: in particular, we introduce the log volatility processes $X^T$ and $X$. In section 3, we give explicit prediction formulas on $X^T$ and $X$. In section 4, we show how to use the results of section 3 to forecast volatility and price options with the MRW model with no knowledge on the value of $\sigma$ or $T$. Finally, in section 5, we give the detailed proof of the results of section 3.

\section{Definitions and preliminaries}
\subsection{Definitions}
For $a<b$, we denote by $\mathcal{D}(]a,b[)$ the space of smooth functions on $]a,b[$ with compact support included in $]a,b[$ and by $\mathcal{D}'(]a,b[)$ the space of distributions on $]a,b[$. 
Let $\mathcal{S}(\R)$ denote the space of Schwartz functions and $\mathcal{S}'(\R)$ it's topological dual  (the space of tempered distributions).  The Fourier transform of an element $f$ of $\mathcal{S}'(\R)$ is defined as usual by:
\begin{equation*}
\forall \phi \in \mathcal{S}(\R), \quad  <\hat{f}, \phi>=<f, \hat{\phi}>,
\end{equation*}
where $\hat{\phi}$ is the classical Fourier transform of $\phi$:
\begin{equation*}
\hat{\phi}(\xi)=\int_{\R}e^{-2i\pi\xi x}\phi(x)dx.
\end{equation*}

We remind the definition of the classical Sobolev Spaces $H^{s}(\R)$ for $s \in \R$: 
\begin{equation*}
H^{s}(\R)=\lbrace f \in  \mathcal{S}'(\R)\; ; \; \hat{f} \in L^{1}_{loc}, \; \int_{\R} (1+|\xi|^{2})^{s}|\hat{f}(\xi)|^2d\xi < \infty  \rbrace.
\end{equation*}
The space $H^{s}(\R)$ equipped with the scalar product:
\begin{equation*}
 <f,g>_{H^{s}}=\int_{\R} (1+|\xi|^{2})^{s}\hat{f}(\xi)\bar{\hat{g}}(\xi)d\xi
\end{equation*}
is a Hilbert space.

We  also introduce the homogeneous Sobolev Spaces $\mathcal{H}^{s}(\R)$ for $s < \frac{1}{2}$: 
\begin{equation*}
\mathcal{H}^{s}(\R)=\lbrace f \in  \mathcal{S}'(\R)\; ; \; \hat{f} \in L^{1}_{loc}, \;  \int_{\R} |\xi|^{2s}|\hat{f}(\xi)|^2d\xi < \infty  \rbrace,
\end{equation*}
The space $\mathcal{H}^{s}(\R)$  equipped with the scalar product:
\begin{equation*}
 <f,g>_{\mathcal{H}^{s}}=\int_{\R} |\xi|^{2s}\hat{f}(\xi)\bar{\hat{g}}(\xi)d\xi
\end{equation*}
is a Hilbert space.

For $s \in [\frac{1}{2};\frac{3}{2}[$, the homogeneous Sobolev Spaces $\mathcal{H}^{s}(\R)$ are defined by:
\begin{equation*}
\mathcal{H}^{s}(\R)=\lbrace f \in  \mathcal{S}'(\R)/ \R \; ;  \;  f' \in \mathcal{H}^{s-1}(\R) \rbrace.
\end{equation*}
For $s \in [\frac{1}{2};\frac{3}{2}[$, the space $\mathcal{H}^{s}(\R)$  equipped with the scalar product:
\begin{equation*}
 <f,g>_{\mathcal{H}^{s}}=\frac{1}{4\pi^2}<f',g'>_{\mathcal{H}^{s-1}}
\end{equation*}
is a Hilbert space.

 \subsection{The processes $(X^{T}_{t})_{t \in \R}$, $(X_{t})_{t \in \R}$ and their reproducing kernel Hilbert space}

We consider the reader is familiar with the theory of gaussian processes. For a complete account on the theory, we refer to \cite{cf:Boga}.  

We introduce the generalized gaussian process $(X^{T}_{t})_{t \in \R}$ 
as a gaussian measure on the space  $\mathcal{S}'(\R)$ with the following characteristic functional:  
\begin{align*}
\forall \phi \in \mathcal{S}(\R), \quad E(e^{i\int_{\R}\phi(t)X^{T}_{t}dt}) & = e^{-1/2E((\int_{\R}\phi(t)X^{T}_{t}dt)^2)} \\ 
& = e^{-1/2\int \int_{\R^2}\phi(t)\phi(s)\ln^{+}(T/|t-s|)dsdt}.
\end{align*}
It follows from Minlos's theorem that the above formula defines a unique gaussian mesure on  $\mathcal{S}'(\R)$ (\cite{cf:GeVi}, chapter 4). Let $\text{sinc}$ be defined by the classical formula:
\begin{equation*}
\text{sinc}(x)=\int_{0}^{x}\frac{\sin(t)}{t}dt.
\end{equation*}
By using Parseval's identity and lemma \ref{lem:app} of the appendix, one can deduce that:
\begin{equation}\label{eq:parseval}
\int \int_{\R^2}\phi(t)\phi(s)\ln^{+}(T/|t-s|)dsdt=\frac{1}{\pi}\int_{\R} \frac{\text{sinc}(2\pi|\xi|T)}{|\xi|} |\hat{\phi}(\xi)|^2d\xi.
\end{equation}

The reproducing kernel Hilbert space $H_{T}$ (\cite{cf:Boga}) associated to $(X^{T}_{t})_{t \in \R}$ is therefore the space:
\begin{equation*}
H_{T}=\lbrace f \in  \mathcal{S}'(\R) \; ;  \;  \int_{\R} \frac{\pi|\xi|}{\text{sinc}(2\pi|\xi|T)} |\hat{f}(\xi)|^2d\xi < \infty  \rbrace.
\end{equation*}
Since there exists two constants $c,C>0$ such that: 
\begin{equation}\label{eq:equivalence}
c(1+|\xi|^2)^{1/2} \leq \frac{|\xi|}{\text{sinc}(2\pi|\xi|)} \leq C(1+|\xi|^2)^{1/2},
\end{equation}
the space  $H_{T}$ is the space ${H}^{1/2}(\R)$ equipped with an equivalent norm.

One would like to let $T$ go to infinity in the above definition. In order to give a rigorous meaning to that procedure, one must work in the space  $\mathcal{S}_{0}(\R)$ of Schwartz functions of average zero:
\begin{equation*}
\mathcal{S}_{0}(\R)=\lbrace   \phi \in  \mathcal{S}(\R) \; ; \; \int_{\R} \phi(t)dt =0  \rbrace.
\end{equation*}
The topological dual of $\mathcal{S}_{0}(\R)$ is then the quotient space  $\mathcal{S}'(\R)/ \R$. We therefore introduce the 
 generalized gaussian process $(X_{t})_{t \in \R}$ 
 as a gaussian measure on the space  $\mathcal{S}'(\R)/ \R$ with the following characteristic function:  
\begin{align}\label{eq:defX}
\forall \phi \in \mathcal{S}_{0}(\R), \quad E(e^{i\int_{\R}\phi(t)X_{t}dt}) & =e^{-1/2E((\int_{\R}\phi(t)X_{t}dt)^2)} \\  
& =e^{-1/2\int \int_{\R^2}\phi(t)\phi(s)\ln(1/|t-s|)dsdt}.
\end{align}

By letting $T$ go to infinity in identity (\ref{eq:parseval}), one gets:
\begin{equation*}
\int \int_{\R^2}\phi(t)\phi(s)\ln(1/|t-s|)dsdt=\frac{1}{2}\int_{\R} \frac{|\hat{\phi}(\xi)|^2}{|\xi|}d\xi.
\end{equation*}
The reproducing kernel Hilbert space $H$ associated to $(X_{t})_{t \in \R}$ is therefore the space:
\begin{equation*}
H=\lbrace f \in  \mathcal{S}'(\R)/ \R \; ;  \;  2 \int_{\R} |\xi| |\hat{f}(\xi)|^2d\xi < \infty  \rbrace.
\end{equation*}
The space $H$ is thus precisely the space $\mathcal{H}^{1/2}(\R)$.

\textbf{Notations}: In the sequel, we will use the notations $k_{T}=\ln^{+}(T/|.|)$ and $k=\ln(1/|.|)$.


\section{The prediction formulas for $X$,$X^{T}$}\label{se:prediction}

\subsection{Prediction formulas for $(X_{t})_{t \in \R}$ }
We have the following explicit formula for the conditional expectation of $(X_{t})_{t \in \R}$:

\begin{theorem}\label{th:pred1}
Let $L$ be some finite positive real number and $f(t)$ the trace on the interval $]-2L,0[$ of a function in 
$\mathcal{H}^{1/2}(\R)$. Then, the conditional expectation  $E[ X_{t} | (X_{s})_{-2L < s < 0}=f ]$ is a function of 
 $\mathcal{H}^{1/2}(\R) $ that satisfies:
  \begin{equation}\label{eq:pred1}
 E[ X_{t} | (X_{s})_{-2L <  s < 0} =f ]=\frac{1}{\pi L} \int_{-2L}^{0}\frac{1-g_{L}(t)^2}{(1-g_{L}(t))^2-2g_{L}(t)\frac{s}{L}}\frac{f(s)ds}{\sqrt{1-(1+\frac{s}{L})^2}},
\end{equation}
where the application $g_{L}$ is given by:
\begin{equation*}
g_{L}(t)=\frac{t}{L}+1-\sqrt{(\frac{t}{L}+1)^2-1}, \quad \text{if} \; t > 0,
\end{equation*}
or 
\begin{equation*}
g_{L}(t)=\frac{t}{L}+1+\sqrt{(\frac{t}{L}+1)^2-1}, \quad \text{if} \; t < -2L.
\end{equation*}
\end{theorem}

\begin{rem}
Since a function in $\mathcal{H}^{1/2}(\R)$ is locally in $L^q$ for all $q<\infty$, we easily check that the integral in \ref{eq:pred1} makes sense.
\end{rem}

By letting $L \to \infty$ in the above formula, we get the following corollary:

\begin{corollary}\label{co:pred1}
Let $f(t)$ be the trace on the interval $]-\infty,0[$  of a function in $\mathcal{H}^{1/2}(\R)$.
 Then, the conditional expectation $E[ X_{t} | (X_{s})_{s < 0} =f]$ is a function of 
 $\mathcal{H}^{1/2}(\R) $ that satisfies:
 \begin{equation}\label{eqco:pred1}
 \forall t > 0 \quad E[ X_{t} | (X_{s})_{s < 0}=f ]=\frac{1}{\pi}\int_{-\infty}^{0}\frac{\sqrt{t}}{(t-s)\sqrt{-s}}f(s)ds.
\end{equation}
\end{corollary}
The proofs of theorem \ref{th:pred1}, corollary \ref{co:pred1} and theorem \ref{th:pred2} (see below) are given in section 5. In the sequel, we will note $K_{L}(t,s)$ the kernel that appears in formula $(\ref{eq:pred1})$ and $K(t,s)$ the kernel that appears in formula $(\ref{eqco:pred1})$.

\begin{rem}\label{rem:def}
Since $(X_{t})_{t \in \R}$ does not belong almost surely to $\mathcal{H}^{1/2}(\R)$, we remind in what sense one must understand formula (\ref{eq:pred1}) (the same remark applies to formula (\ref{eqco:pred1})). An equivalent formulation of (\ref{eq:pred1}) is for all $\phi$ in $\mathcal{S}_{0}(\R)$ with support in $]-\infty,-2L[ \cup ]0,\infty[$ :
\begin{equation*}
E \big[ \int_{\R}\phi(t)X_{t}dt \; | \left( \int_{-2L}^{0}\psi(s)X_{s}ds \right)_{\psi \in \mathcal{D}(]-2L,0[);\int \psi=0} \big] = \int_{-2L}^{0} \left( \int_{\R}\phi(t)K_{L}(t,s)dt \right) X_{s}ds
\end{equation*}

\end{rem}

\subsection{Consequence on the prediction of $(X^{T}_{t})_{t \in \R}$}

As a consequence of the above formulas on $(X_{t})_{t \in \R}$, it is possible to obtain an exact prediction formula on  $(X^{T}_{t})_{t \in \R}$.\textbf{This formula is no longer defined modulo constants and is valid whatever the value of the correlation length $T$}. It is obtained as a perturbation of the case $T=\infty$.  

In order to state the formula, we introduce $\varphi$ as the unique distribution of $H^{-1/2}(\R)$ with support in $[-1,0]$ solution of:
\begin{equation*}
\forall t \in ]0,1[, \quad \int_{-1}^{0}\ln(1/|t-s|)\varphi(s)ds=1.
\end{equation*}
We refer to section 5.3 for the existence and uniqueness of $\varphi$. In fact, the distribution $\varphi$ is a function and it is given explicitly by the following formula:
\begin{equation}\label{eq:explicit}
\forall s \in ]-1,0[, \quad \varphi(s)=\frac{1}{2\pi\ln(2)}\frac{1}{\sqrt{-s-s^2}}
\end{equation}

 This leads to:

\begin{theorem}\label{th:pred2}
Let $L$ be some finite and positive real number such that $2L<T$ and $f(t)$ the trace on $]-2L,0[$ of a function in $H^{1/2}(\R)$. Then $E[ X^{T}_{t} | (X^{T}_{s})_{-2L < s < 0}=f ]$ is a function of $H^{1/2}(\R)$ that satisfies:
 \begin{equation}\label{eq:pred2}
 \forall t \in ]0,T-2L[, \quad  E[ X^{T}_{t} | (X^{T}_{s})_{-2L < s < 0}=f ]= \int_{-2L}^{0}K_{L,T}(t,s)f(s)ds,
\end{equation}
where the kernel satisfies for $t$ in $]0,T-2L[$:
\begin{equation}\label{eq:kernel}
K_{L,T}(t,s)=K_{L}(t,s)+\frac{(k*\varphi)(t/2L)-1}{1+\frac{\ln(T/2L)}{2\ln(2)}}\frac{1}{\pi\sqrt{-s(2L+s)}}.
\end{equation}
\end{theorem}

A remark similar to remark \ref{rem:def} applies to formula (\ref{eq:pred2}).

\section{Application to volatility forecasting and option pricing}

\textbf{In this section, we suppose that we ignore the values of $\sigma$ and $T$ but that the value of $\lambda^2$ is known.}

\subsection{Forecasting volatility}

\subsubsection{Discretizing formula (\ref{eqco:pred1})}
Let us first suppose that $T$ is large compared to $L$ so that the prediction kernel is given by $K_{L}$. Suppose also that the time $t$ at which we intend to predict is small compared to $L$. It follows that we may use the prediction formula (\ref{eqco:pred1}):
\begin{equation*}
E[ X_{t} | (X_{s})_{s < 0}=f ]=\frac{1}{\pi}\int_{-\infty}^{0}\frac{\sqrt{t}}{(t-s)\sqrt{-s}}f(s)ds.
\end{equation*}
Let us take some small time lag $\tau$ and discretize the above formula. We get:
\begin{equation}\label{eq:disc}
f \simeq \sum_{k=0}^{N}1_{[-(k+1)\tau,-k\tau[}f(-k\tau)
\end{equation}
and thus for $n \geq 1$:
\begin{equation*}
E[ X_{n \tau} | (X_{-k\tau})_{0 \leq k \leq N}=(f(-k\tau))_{0 \leq k \leq N}] \simeq \frac{1}{\pi} \sum_{k=0}^{N} \int_{-(k+1)\tau}^{-k\tau}\frac{\sqrt{n\tau}}{(n\tau-s)\sqrt{-s}}dsf(-k\tau).
\end{equation*}
Let us denote 
\begin{align*}
 \alpha_{n,k}^{*} & = \frac{1}{ \pi} \int_{(k-1)/n}^{k/n}\frac{ds}{(1+s)\sqrt{s}} \\
 & =\frac{2}  {\pi}(\text{Arctan} \sqrt{\frac{k}{n}}-\text{Arctan} \sqrt{\frac{k-1}{n}}),
 \end{align*}
so that the discretized formula writes:
\begin{equation*}
E[ X_{n \tau} | (X_{-k\tau})_{0 \leq k \leq N}] \simeq  \sum_{k=0}^{N} \alpha_{n,k+1}^{*}X_{-k\tau}. 
\end{equation*}
Of course, we have for all $n \geq 1$, $\sum_{k=0}^{N} \alpha_{n,k+1}^{*}=\frac{2}{\pi}\text{Arctan}\sqrt{\frac{N}{n}}$ so $\sum_{k=0}^{N} \alpha_{n,k+1}^{*} \simeq 1$ for $n \ll N$. Notice also that $ \alpha_{n,k}^{*}$ does not depend on $\tau$. 

\subsubsection{Application of the discretized formula}

We will model the (log) return $(r_{n})_{n \in \Z}$ at scale $\tau$ of a financial asset as a discretized MRW:
\begin{equation*}
r_{n}=\sigma_{n}\epsilon_{n}
\end{equation*}
where $(\epsilon_{n})_{n \in \Z}$ is the noise process (i.i.d. standard gaussian) and $(\sigma_{n})_{n \in \Z}$ the volatility process:
\begin{equation*}
\sigma_{n}=\sigma \sqrt{\tau}e^{\lambda X_{n}^{\tau}-\lambda^2\ln(T / \tau)}
\end{equation*}
with $X_{n}^{\tau}$ given by definition (\ref{eq:defiX}).

Suppose one can observe the historical volatility $\sigma_{n}$ (or equivalently $X_{n}^{\tau}$) on some time window $\{-N,...,0\}$ (Of course this is rigorously not the case in finance and one must use filtering theory, intraday data, etc... to get a proxy of $\sigma_{n}$).
We can decompose the process $X_{n}^{\tau}$, $n \geq 1$:
\begin{equation*}  
X_{n}^{\tau}= \sum_{k=0}^{N} \alpha_{n,k+1}X_{-k}^{\tau}+Z_{n} \quad (\ast), 
\end{equation*}
where $Z_{n}$ is a centered gaussian process independent from $X_{-N}^{\tau}, \ldots, X_{0}^{\tau}$ and the coefficients 
$\alpha_{n,k+1}$ are uniquely determined by the conditions:
\begin{equation*}  
 E[X_{j}^{\tau}Z_{n}]=0, \quad j=-N,\ldots,0. 
\end{equation*}
If we suppose that $N \ll T/ \tau$ and $n \ll N$ then formula $(\ast)$ appears as a discretization of formula (\ref{eqco:pred1}) and we may approximate:
\begin{equation*}  
\alpha_{n,k+1} \simeq \alpha_{n,k+1}^{*}
\end{equation*}
so that $\sum_{k=0}^{N}\alpha_{n,k+1} \simeq 1$. Then we write:
\begin{align*}  
\sigma_{n} & \simeq e^{\lambda Z_{n}} \prod_{k=0}^{N}(\sigma \sqrt{\tau}e^{\lambda X_{-k}^{\tau}-\lambda^2\ln(T / \tau)})^{\alpha_{n,k+1}^{*}}  \\
& \simeq e^{\lambda Z_{n}} \prod_{k=0}^{N}(\sigma_{-k})^{\alpha_{n,k+1}^{*}},
\end{align*}
from which we get the conditional expectation:
\begin{equation*}
E[\sigma_{n} | (\sigma_{-k})_{0 \leq k \leq N}] \simeq e^{\frac{\lambda^2}{2} E[Z_{n}^2]} \prod_{k=0}^{N}(\sigma_{-k})^{\alpha_{n,k+1}^{*}}.
\end{equation*}
Now let us calculate $E[Z_{n}^2]$.
We use:
\begin{equation*}
E[(X_{n}^{\tau})^2] \simeq E[(\sum_{k=0}^{N}\alpha_{n,k+1}^{*}X_{-k}^{\tau})^2]+E[Z_{n}^2]
\end{equation*}
and
\begin{equation*}
\sum_{k=0}^{N}\alpha_{n,k+1}^{*}X_{-k}^{\tau} \simeq \frac{1}{\pi} \int_{-\infty}^{0}\frac{\sqrt{n\tau}}{(n\tau-s)\sqrt{-s}}X_{s}ds.
\end{equation*}
Therefore, we get:
\begin{equation*}
E[Z_{n}^2]=\ln(n)+\frac{1}{\pi^2} \int_{0}^{\infty}\int_{0}^{\infty}\frac{1}{(1+s)(1+\tilde{s})\sqrt{s}\sqrt{\tilde{s}}}\ln(|s-\tilde{s}|)dsd\tilde{s}.
\end{equation*}
If we set $C$ equal to the constant term in the above expression, we get the following simple formula:
\begin{equation*}
E[\sigma_{n} | (\sigma_{-k})_{0 \leq k \leq N}] \simeq e^{\frac{\lambda^2C}{2}}n^{\lambda^2/2} \prod_{k=0}^{N}(\sigma_{-k})^{\alpha_{n,k+1}^{*}}.
\end{equation*}
One can notice the remarkable fact that, in the above formula, $\sigma,\tau,T$ have disappeared. Nevertheless, the formula depends on $\lambda^2$ and we get the following sensitivity with respect to $\lambda^2$:
\begin{equation}\label{eq:der}
\frac{\delta E[\sigma_{n} | (\sigma_{-k})_{0 \leq k \leq N}]}{\delta \lambda^2}=(\frac{C}{2}+\frac{\ln(n)}{2})E[\sigma_{n} | (\sigma_{-k})_{0 \leq k \leq N}].
\end{equation}
Since $C \simeq 1.33$ numerically, if one considers the problem of forecasting volatility on a period of $1$ month ($n \leq 20$) and $\lambda^2$ is known with a precision $\delta \lambda^2 =0.01$, the above formula (\ref{eq:der}) implies that the forecast is correct with a very high precision of  approximately $2$ percent. 

%


\subsection{Option pricing with unknown parameters}
Let $S_{t}$ be the price process of a financial asset. We model the log price by the continuous MRW $Y^{T}$ (with underlying parameters $(\sigma, \lambda,T)$) given by expression (\ref{eq:defMRW}). 
In this subsection, we fix a real positive number $t$. Suppose one can observe the historical (log) volatility $X_{s}^{T}$ on some time window $]-2L,0[$. The price of a standard call option with maturity $t$ and exercise price $K$ is then given by (for simplicity, we suppose the risk free rate $r=0$ and that there are no dividends):
\begin{equation}\label{eq:call}
E [(S_{0}e^{Y_{t}^{T}}-K)_{+}) | (X^{T}_{s})_{-2L < s < 0}=f],
\end{equation}
where $Y_{t}^{T} $ is given by the approximation (\ref{eq:approximation}). For simplicity, we will also suppose that the maturity $t$ is not very large (typically, a few months) such that one can make the following approximation by an additive model:
\begin{equation*}
S_{0}e^{Y_{t}^{T}} \simeq S_{0}(1+Y_{t}^{T}).
\end{equation*}
Then, one can perform a cumulant expansion in formula (\ref{eq:call}) up to the kurtosis (see p. 244-245 in \cite{cf:BoPo} for details) and obtain the following smile formula for the implied volatility $\Sigma(K,t)$ as a function of the strike $K$ and the maturity $t$:
\begin{equation}\label{eq:smile}
\Sigma(K,t)=\frac{\sigma_{t}}{\sqrt{t}}\big(1+\kappa_{t}(\frac{(K-S_{0})^2}{S_{0}^2  \sigma_{t}^2}-1)\big),
\end{equation}
where $\sigma_{t}^2$ is the forecasted variance:
\begin{equation}\label{eq: volforecast}
\sigma_{t}^2=E[(Y_{t}^{T})^2 | (X^{T}_{s})_{-2L < s < 0}=f]
\end{equation}
and $\kappa_{t}$ the forecasted kurtosis:
\begin{equation}\label{eq: kurforecast}
\kappa_{t}=3(\frac{E[(Y_{t}^{T})^4 | (X^{T}_{s})_{-2L < s < 0}=f]}{E[(Y_{t}^{T})^2 | (X^{T}_{s})_{-2L < s < 0}=f]^2}-1).
\end{equation}
We outline the computation of (\ref{eq: volforecast}). Let $\tau$ be some small observation scale and let $r_{n}=Y_{\tau n}^{T}-Y_{\tau(n-1)}^{T}$ be the (log) return at scale $\tau$. We have:
\begin{equation*}
\sigma_{t}^2=\sum_{n=1}^{\lfloor t / \tau \rfloor }E[r_{n}^2 | (X^{T}_{s})_{-2L < s < 0}=f]
\end{equation*}
One then uses the expression $r_{n}=\sigma  \sqrt{\tau}e^{\lambda X_{n}^{\tau}-\lambda^2\ln(T / \tau)} \epsilon_{n}$ of the previous subsection and the discretization:
\begin{equation*}
f \simeq \sum_{k=0}^{N}1_{[-(k+1)\tau,-k\tau[}f(-k\tau).
\end{equation*}
We therefore get with $N= \lfloor 2L / \tau  \rfloor$:
\begin{equation*}
\sigma_{t}^2 \simeq \sum_{n=1}^{\lfloor t / \tau \rfloor }E[r_{n}^2 | (X_{-k}^{\tau})_{0 \leq k \leq N}=(f(-k\tau))_{0 \leq k \leq N}].
\end{equation*}
The computation of each term 
\begin{equation*}
E[r_{n}^2 | (X_{-k}^{\tau})_{0 \leq k \leq N}=(f(-k\tau))_{0 \leq k \leq N}]=E[\sigma^2 \tau e^{2\lambda X_{n}^{\tau}-2\lambda^2\ln(T / \tau)} | (X_{-k}^{\tau})_{0 \leq k \leq N}=(f(-k\tau))_{0 \leq k \leq N}]
\end{equation*} 
is similar to the computations of the previous subsection.

Since, in practice, $\lambda^2$ is roughly found around the value $0.02$, we perform an expansion of  $Y_{t}^{T}$ for $\lambda^2 \ll 1$ in order to compute $\kappa_{t}$.
We have the following scaling identity:
\begin{equation*}
Y_{t}^{T}\underset{(Law)}{=}\sigma\sqrt{\theta(t)}\epsilon,
\end{equation*}
where $\epsilon$ is a standard gaussian variable independent of $\theta(t)$. 
We can derive formally the following series of approximations for $\lambda^2 \ll 1$ (we will use for $c^2 \ll 1$ and $X$ a generalized centered gaussian variable the approximation $e^{cX-c^2E(X^2)/2} \simeq 1+cX$: see \cite{cf:BaKoMu} for an exact mathematical formulation):
\begin{align}
Y_{t}^{T} & =\sigma\sqrt{\theta(t)}\epsilon \nonumber \\
&=  \sigma \sqrt{\int_{0}^{t}e^{2\lambda X_{s}^{T}-2\lambda^2E((X_{s}^{T})^2)}ds} \; \epsilon 
\nonumber \\ 
&=   \sigma  \sqrt{t} \sqrt{\frac{1}{t}\int_{0}^{t}e^{2\lambda X_{s}^{T}-2\lambda^2E((X_{s}^{T})^2)}ds} \;\epsilon  \nonumber \\
& \simeq  \sigma \sqrt{t} \sqrt{\frac{1}{t}\int_{0}^{t}(1+2\lambda X_{s}^{T})ds} \; \epsilon  \nonumber \\
&=   \sigma  \sqrt{t} \sqrt{1+\frac{ 2 \lambda}{t}\int_{0}^{t}X_{s}^{T}ds} \; \epsilon \nonumber \\
& \simeq  \sigma \sqrt{t} e^{\frac{\lambda}{t}\int_{0}^{t}X_{s}^{T}ds-\lambda^2 \ln(\frac{Te^{3/2}}{t})} \epsilon \label{eq:approximation}  
\end{align}
where $\epsilon$ is a standard gaussian variable independent of $X^{T}$. 
Using expression (\ref{eq:approximation}), standard computations give the following 
expression for $\kappa_{t}$:
\begin{equation}\label{eq:kurt1}
\kappa_{t}=3(e^{\frac{4\lambda^2}{t^2}E[(\int_{0}^{t}X_{s}^{T}ds-\int_{[0,t] \times [-2L,0]}K_{L,T}(s,\sigma)X_{\sigma}^{T}d\sigma ds)^2]}-1).
\end{equation}
If we let $T$ go to infinity, we get the following limit expression:
\begin{equation}\label{eq:kurt2}
\kappa_{t} \underset{T \to \infty}{\rightarrow}3(e^{\frac{4\lambda^2}{t^2}\int_{[0,t]^2 \times [-2L,0]^2}K_{L}(s,\sigma)K_{L}(\tilde{s},\tilde{\sigma})\ln(\frac{|s-\tilde{\sigma}| |\tilde{s}-\sigma|}{|s-\tilde{s}| |\sigma-\tilde{\sigma}|})ds d\tilde{s} d\sigma d\tilde{\sigma}}-1). 
\end{equation}
\begin{rem}
The formulas (\ref{eq:kurt1}) and (\ref{eq:kurt2}) are independant of $f$, which is a consequence of the lognormal assumption: every lognormal model for the volatility will give a constant conditional kurtosis. Intuitively, this can appear as a limitation of the model.   
\end{rem}
\begin{rem}
It is remarkable that the conditional kurtosis given by formula (\ref{eq:kurt1}) tends to a finite limit given by (\ref{eq:kurt2}) as $T$ goes to infinity. Indeed, the unconditional kurtosis tends to infinity as $T$ goes to infinity.
\end{rem}

\section{Proof of the prediction formulas}

\subsection{Preliminary results and definition}

In this section, we prove the theorems of section \ref{se:prediction}.
Let $\Pi$ denote the upper half plane $\R \times ]0;+\infty[$ and  $V^{1}(\Pi)$ denote the following quotient space:
 \begin{equation*}
 V^{1}(\Pi)=\lbrace  u \in \mathcal{D}'(\Pi)/ \R \; ;  \; \int_{\Pi}| \nabla u |^2 dxdy < \infty \rbrace
 \end{equation*} 
equipped with the norm 
\begin{equation*}
|u|_{ V^{1}(\Pi)}= \sqrt{ \int_{\Pi} | \nabla u |^2  dxdy}.
\end{equation*}

On the Hilbert space $V^{1}(\Pi)$, one can define in a classical way a trace operator:

\begin{lemma}   
There exists  a unique continuous linear operator $\Gamma: V^{1}(\Pi) \rightarrow \mathcal{H}^{1/2}(\R)$ that satisfies:
 \begin{equation*}
\forall f \in \mathcal{S}(\R^2), \quad \Gamma(f_{|y>0})(x)=f(x,0).
\end{equation*} 
The operator $\Gamma$ is onto and satisfies:
\begin{equation*}
|\Gamma(f)|_{\mathcal{H}^{1/2}(\R)}^2=\frac{1}{2\pi}|f|_{ V^{1}(\Pi)}^2.
\end{equation*}

\end{lemma}



We introduce the Poisson kernel $P$ by the following formula: 
 \begin{equation}\label{eq:poisson}
\forall f \in  \mathcal{H}^{1/2}(\R), \quad P(f)(x,y)=\frac{1}{\pi}\int_{\R}\frac{|y|}{y^2+(x-z)^2}f(z)dz.
\end{equation}  
Notice that the existence of the above integral is not obvious: one first defines it for $f$ in $\mathcal{S}(\R)$ and then extends it to $\mathcal{H}^{1/2}(\R)$ by density.  

The next lemma relates $P$ to a variational problem.

\begin{lemma}   
For all $f \in \mathcal{H}^{1/2}(\R)$ there exists a unique solution to the variational problem:
 \begin{equation}\label{eq:mini}
\inf_{u \in V^{1}(\Pi) \; ; \; \Gamma(u)=f}(\int_{\Pi} | \nabla u|^2 dxdy).
\end{equation} 
and the solution is $P(f)$ given by the Poisson formula \ref{eq:poisson}.
Moreover, $P(f)$ is the unique solution to the Dirichlet problem:
\begin{equation}\label{eq:Dipro}
\begin{cases}
u \in V^{1}(\Pi)  \\
\Delta u=0                            \\
\Gamma(u)=f
\end{cases}
\end{equation}
and the following identity holds:
\begin{equation}\label{eq:isometry}
|P(f)|_{ V^{1}(\Pi)}^2= 2\pi |f|_{ \mathcal{H}^{1/2}(\R)}^2.
\end{equation}
\end{lemma}

\proof

Let $f \in \mathcal{H}^{1/2}(\R)$ be fixed.
A simple computation shows that the $\R^2$ Fourier transform of $P(f)$ is given by:
 \begin{equation*}
\hat{P(f)}(\xi_{1},\xi_{2})=\frac{1}{\pi}\hat{f}(\xi_{1})\frac{|\xi_{1}|}{\xi_{1}^2+\xi_{2}^2}.
\end{equation*}
Therefore one gets:
\begin{align*}
|P(f)|_{V^{1}(\Pi)}^2 & = \frac{1}{2}\int_{\R^2}| \nabla P(f)|^2 dxdy\\
& = 2 \int_{\R^2}|\hat{f}(\xi_{1})|^2\frac{\xi_{1}^2}{\xi_{1}^2+\xi_{2}^2} d\xi_{1}d\xi_{2} \\
& = 2\pi  \int_{\R}|\xi||\hat{f}(\xi)|^2 d\xi,
\end{align*}
 which is precisely identity \ref{eq:isometry}. It is straightforward to check that $P(f)$ is solution to \ref{eq:Dipro}. 
One can easily show that $\mathcal{D}(\Pi)$ is dense in $Ker(\Gamma)$ which implies that:   
 \begin{equation*}
\forall u \in Ker(\Gamma), \quad \int_{\Pi} (\nabla P(f).  \nabla u) dxdy=0;
\end{equation*}  
and therefore, one gets for all $u$ in $Ker(\Gamma)$ the following identity:
 \begin{equation*}
\int_{\Pi} | \nabla (P(f)+u) |^2 dxdy=\int_{\Pi} | \nabla P(f) |^2 dxdy+\int_{\Pi} | \nabla u |^2 dxdy.
\end{equation*}  
This entails that $P(f)$ is the unique minimizer of \ref{eq:mini}. 

\qed

\begin{corollary}\label{co:DiproNeu}
Let $f$ be some element of $\mathcal{H}^{1/2}(\R)$ and $]a,b[$ an open interval . Consider the function $m=f_{|]a,b[}$ and the associated variational problem:  
 \begin{equation*}
\inf_{g \in \mathcal{H}^{1/2}(\R) \; ; \; g_{|]a,b[}=m}|g|_{ \mathcal{H}^{1/2}(\R)}^2.
\end{equation*}
Suppose $u$ is a solution of the following Dirichlet problem:
\begin{equation}\label{eq:DiproNeu}
\begin{cases}
u \in V^{1}(\Pi)  \\
\Delta u=0                            \\
\frac{\partial u}{\partial n}_{|\R \setminus ]a,b[}=0 \\
\Gamma(u)_{|]a,b[}=m
\end{cases}
\end{equation}
Then $\Gamma(u)$ is the unique solution to the above variational problem. 
\end{corollary}
\proof
Let g be an element of  $\mathcal{H}^{1/2}(\R)$ such that $g_{|]a,b[}=m$. Then we have the following identity:

\begin{align*}
|g|_{\mathcal{H}^{1/2}(\R)}^2  & =\frac{1}{2\pi} \int_{\Pi} | \nabla (P(g)) |^2 dxdy  \\
&= \frac{1}{2\pi} \int_{\Pi} | \nabla (u) |^2 dxdy+ \frac{1}{\pi} \int_{\Pi} (\nabla (P(g)-u). \nabla u) dxdy+\frac{1}{2\pi} \int_{\Pi} | \nabla (P(g)-u) |^2 dxdy\\
& \underset{(Green)}{\geq} |\Gamma(u)|_{ \mathcal{H}^{1/2}(\R)}^2 +\underset{=0}{\underbrace{\int_{\R}(g(x)-\Gamma(u)(x))\frac{\partial u}{\partial n}(x,0)dx}} \\
& \geq |\Gamma(u)|_{ \mathcal{H}^{1/2}(\R)}^2.
\end{align*}

\subsection{Proof of theorem \ref{th:pred1}}
Let $f$ be the trace on $]-2L,0[$ of a $\mathcal{H}^{1/2}(\R)$ function. By a general property of gaussian measures (\cite{cf:Wah}), 
the conditional expectation on the left hand side can be obtained as the solution of the variational problem:
 \begin{equation*}
\inf_{g \in \mathcal{H}^{1/2}(\R) \; ; \; g_{|]-2L,0[}=f}|g|_{ \mathcal{H}^{1/2}(\R)}^2.
\end{equation*}
By corollary \ref{co:DiproNeu}, one must solve the Dirichlet problem \ref{eq:DiproNeu}:
\begin{equation*}
\begin{cases}
u \in V^{1}(\Pi)  \\
\Delta u=0                            \\
\frac{\partial u}{\partial n}_{|\R \setminus ]-2L,0[}=0 \\
\Gamma(u)_{|]-2L,0[}=f
\end{cases}
\end{equation*}
The application $u$ is solution to the above Dirichlet problem if and only if $U$ given by $U(x,y)=u(Lx-L,Ly)$ is solution to:
\begin{equation*}
\begin{cases}
U \in V^{1}(\Pi)  \\
\Delta U=0                            \\
\frac{\partial U}{\partial n}_{|\R \setminus ]-1,1[}=0 \\
\Gamma(U)_{|]-1,1[}=M
\end{cases}
\end{equation*}
where $M(x)=f(Lx-L)$. 
One can solve this problem explicitly by using conformal mappings.
Let $\phi$ be the conformal map from the half disc 
\begin{equation*}
\mathbb{D}_{+}=\lbrace z \in \mathbb{C}; \; |z|<1 \; \text{and} \; Re(z) > 0 \rbrace
\end{equation*}
 to the upper half plane $\Pi$: 
\begin{equation*}
\phi(z)=\frac{1}{2i}(z-\frac{1}{z}).
\end{equation*}
Then $v(z) = U(\phi(z))$ is harmonic in $\mathbb{D}_{+}$, satisfies the condition $\frac{\partial v}{\partial n}=0$ on the vertical diameter $\lbrace Re(z) =0 \rbrace \cap \partial \mathbb{D}_{+}$ and we have the following equality for $x^2+y^2=1$:
\begin{equation*}
v(x,y)=U(y,0)=M(y).
\end{equation*}
Since $\frac{\partial v}{\partial n}=0$ on the vertical diameter, one can extend symmetrically $v$ as a harmonic function on the unit disc. By Poisson's formula, $v$ has the following representation:
\begin{equation*}
v(re^{i\theta})=\frac{1}{2\pi}\int_{-\pi}^{\pi}\frac{1-r^2}{1-2r\cos(\theta-t)+r^2}v(e^{it})dt.
\end{equation*}
In particular, one gets:
\begin{equation*}
 v(0,y)=\frac{1}{\pi}\int_{-\pi /2 }^{\pi/2}\frac{1-y^2}{1-2y\sin(t)+y^2}M(sin(t))dt.
\end{equation*}
For $X$ greater or equal to $1$,
\begin{equation*}
U(X,0)=v(\phi^{-1}(X,0))=v(0,X-\sqrt{X^2-1}).
\end{equation*}
and for $X$ less or equal to $-1$,
\begin{equation*}
U(X,0)=v(0,X+\sqrt{X^2-1}).
\end{equation*}

Therefore, one gets
\begin{equation*}
u(x,0)= \frac{1}{\pi}\int_{-\pi /2 }^{\pi/2}\frac{1-g_{L}(x)^2}{1-2g_{L}(x)\sin(t)+g_{L}(x)^2}f(L(\sin(t)-1))dt,
\end{equation*}
where $g_{L}$ is given by the formula:
\begin{equation*}
g_{L}(x)=\frac{x}{L}+1-\sqrt{(\frac{x}{L}+1)^2-1}, \quad \text{if} \; x > 0,
\end{equation*}
or 
\begin{equation*}
g_{L}(x)=\frac{x}{L}+1+\sqrt{(\frac{x}{L}+1)^2-1}, \quad \text{if} \; x < -2L,
\end{equation*}
One gets the desired result by making the change of variable $s=L(\sin(t)-1)$ in the above integral.

\qed 

\subsection{Proof of corollary \ref{co:pred1}}
We take the limit $L \rightarrow \infty$ in formula (\ref{eq:pred1}). It is straightforward to check the convergence in 
$\mathcal{D}'(]0,\infty[)$ of the right hand term. Let us denote $w_{L}^{*}$ the left hand term in (\ref{eq:pred1}) and $w^{*}$ the left hand term in (\ref{eqco:pred1}). We have:
\begin{equation*}
|w_{L}^{*}|_{{\mathcal{H}}^{1/2}(\R)}=\inf_{w \in {\mathcal{H}}^{1/2}(\R) \; ; \; w_{|]-2L,0[}=f}|w|_{{\mathcal{H}}^{1/2}(\R)}
\end{equation*}  
and
\begin{equation*}
|w^{*}|_{{\mathcal{H}}^{1/2}(\R)}=\inf_{w \in {\mathcal{H}}^{1/2}(\R) \; ; \; w_{|]-\infty,0[}=f}|w|_{{\mathcal{H}}^{1/2}(\R)}.
\end{equation*} 
Thus we have $|w_{L}^{*}|_{{\mathcal{H}}^{1/2}(\R)} \leq |w^{*}|_{{\mathcal{H}}^{1/2}(\R)}$. Modulo a subsequence, we may suppose by the Banach-Alaoglu theorem that $w_{L}^{*}$ converges weakly in ${\mathcal{H}}^{1/2}(\R)$ as $L$ goes to infinity to $w^{**}$.  We have the convergence of $w_{L}^{*}$ to $w^{**}$ in $\mathcal{D}' \setminus \R$ which implies that $w_{|]-\infty,0[}^{**}=f$. Since $|w^{**}|_{{\mathcal{H}}^{1/2}(\R)} \leq |w^{*}|_{{\mathcal{H}}^{1/2}(\R)}$, we get $w^{**}=w^{*}$.

\subsection{Proof of theorem \ref{th:pred2}}

\subsubsection{The minimization problems}
We consider the Sobolev space  ${H}^{1/2}(\R)$ equipped with the norm $|.|_{T}$ given by $H_{T}$:
\begin{equation*}
|v|_{T}^2=\int_{\R}\frac{1}{\hat{k}_{T}(\xi)}|\hat{v}(\xi)|^2d\xi,
\end{equation*}
where $\hat{k}_{T}$ is the Fourier transform of ${k}_{T}$:
\begin{equation*}
\hat{k}_{T}(\xi)=\frac{1}{\pi|\xi|}\text{sinc}(2\pi|\xi|T).
\end{equation*}
Let $f$ be some function of $]-2L,0[$ which is the restriction of some function of  ${H}^{1/2}(\R)$. We will denote by $v^{*}_{L,T}$ the function of ${H}^{1/2}(\R)$ solution of the minimization problem:
\begin{equation*}
\inf_{v \in {H}^{1/2}(\R) \; ; \; v_{|]-2L,0[}=f}|v|_{T}^2  \quad  \quad  \quad (V_{T})
\end{equation*}
We can consider the function $f$ as a function of $\mathcal{H}^{1/2}(\R)$ and note $v^{*}_{L}$ the unique solution of the minimization problem:
\begin{equation*}
\inf_{v \in {\mathcal{H}}^{1/2}(\R) \; ; \; v_{|]-2L,0[}=f}|v|_{{\mathcal{H}}^{1/2}(\R)}^2 \quad \quad \quad (V_{\infty})
\end{equation*}

\subsubsection{A few intermediate propositions}
We will start by giving a few intermediate propositions before proving the theorem \ref{th:pred2}.
\begin{proposition}\label{prop:exist}
Let $f$ be some function of ${H}^{1/2}(\R)$.There exists a unique $\alpha_{L,T}$ in ${H}^{-1/2}(\R)$ with support in $[-2L,0]$ that satisfies:
\begin{equation}\label{eq:conv}
\forall t \in ]-2L,0[, \quad (k_{T} \ast \alpha_{L,T})(t)= f(t). 
\end{equation}
\end{proposition}
\proof
Consider the Hilbert space $V$ of distributions in $H^{-1/2}(\R)$ with support in $[-2L,0]$. On the product $V \times V$, we consider the bilinear form:
\begin{align*}
a(u,v) & = \int_{\R}(k_{T} \ast u)(t)v(t)dt\\
& = \int_{\R}\hat{k}_{T}(\xi)\hat{u}(\xi)\bar{\hat{v}}(\xi)d\xi.\\
\end{align*}
By inequality (\ref{eq:equivalence}),  $\sqrt{a(u,u)}$ defines a norm equivalent to ${H}^{-1/2}(\R)$. The linear form $\mathcal{L}: v \rightarrow \int_{\R}f(t)v(t)dt$ is continuous on $V$. By the Lax-Milgram theorem, there exists a unique $\alpha_{L,T}\in V$ such that:
\begin{equation*}
\forall v \in V, \quad a(\alpha_{L,T},v)=\mathcal{L}(v). 
\end{equation*} 
Since $\mathcal{D}(]-2L,0[) \subset V$, we get the desired equality.
\qed

In the sequel, we will denote by $\varphi_{L,T}$ the solution to equation (\ref{eq:conv}) with $f=1$ and  $\varphi=\varphi_{1/2,1}$.
 Note that one can show the obvious scaling equality when $2L \leq T$:
\begin{equation}\label{eq:scale}
\varphi_{L,T}(t)=\frac{1}{2L}\frac{\varphi(t/2L)}{1+\ln(T/2L)\int_{-1}^{0} \varphi(s)ds},
\end{equation}
with $\varphi$ given by (cf.appendix): 
\begin{equation*}
\forall s \in ]-1,0[, \quad \varphi(s)=\frac{1}{2\pi\ln(2)}\frac{1}{\sqrt{-s-s^2}}.
\end{equation*}
We have thus the following expression for $\varphi_{L,T}$:
\begin{equation*}
\varphi_{L,T}(t)=\frac{1}{2\pi\ln(2)+\pi \ln(T/2L)}\frac{1}{\sqrt{-s(s+2L)}}.
\end{equation*}

One can apply the above proposition to the resolution of the minimization problems $(V_{T})$ and 
$(V_{\infty})$.

\begin{proposition}
Let $\alpha_{L,T}$ be the solution of the equation (\ref{eq:conv}). Then the function $v^{*}_{L,T}=k_{T} \ast \alpha_{L,T}$ is solution to $(V_{T})$ .
\end{proposition}
\proof
Let $v \in H^{1/2}(\R)$ be such that $v=f$ on $]-2L,0[$. We get:
\begin{align*}
|v^{*}_{L,T}|_{T}^{2} & = \int_{\R}\hat{k}_{T}(\xi)|\hat{\alpha}_{L,T}(\xi)|^2d\xi   \\
& = \int_{\R}v^{*}_{L,T}(t)\alpha_{L,T}(t)dt \\
& = \int_{\R}v(t)\alpha_{L,T}(t)dt \\
& = \int_{\R}\hat{v}(\xi)\bar{\hat{\alpha}}_{L,T}(\xi)d\xi \\
& \leq |v|_{T}|v^{*}_{L,T}|_{T}. \\
\end{align*}
\qed

Similarly, wet get the following solution to  $(V_{\infty})$:
\begin{proposition}
Let $f$ be in $\mathcal{H}^{1/2}(\R)$. Let $\alpha_{L}$ be the solution in ${H}^{-1/2}(\R)$ with support in 
$[-2L,0]$ of the equation: 
\begin{equation*}
\forall t \in ]-2L,0[, \quad (k_{2L} \ast \alpha_{L})(t)= f(t)-\frac{\int_{-2L}^{0}f(s)\varphi(s/2L)ds}{\int_{-2L}^{0}\varphi(s/2L)ds}.
\end{equation*}
Then $v_{L}^{*}= k \ast \alpha_{L}+\frac{\int_{-2L}^{0}f(s)\varphi(s/2L)ds}{\int_{-2L}^{0}\varphi(s/2L)ds}$ is solution to $(V_{\infty})$. 
\end{proposition}
\proof
First note that by the scaling relation (\ref{eq:scale}), we have:
\begin{equation*}
\frac{\int_{-2L}^{0}f(s)\varphi(s/2L)ds}{\int_{-2L}^{0}\varphi(s/2L)ds}=\frac{\int_{-2L}^{0}f(s)\varphi_{L,2L}(s)ds}{\int_{-2L}^{0}\varphi_{L,2L}(s)ds}
\end{equation*}
Thus we get that:
\begin{align*}
\int_{\R}\alpha_{L}(t)dt & = \int_{\R}(k_{2L} \ast \varphi_{L,2L})(t)\alpha_{L}(t)dt \\
&=  \int_{\R}(k_{2L} \ast \alpha_{L})(t)\varphi_{L,2L}(t)dt  \\
&=0. \\
\end{align*}
It is obvious by definition that $v^{*}_{L}$ is in $\mathcal{H}^{1/2}(\R)$ and that $v^{*}_{L}=f$ on the interval $]-2L,0[$.
Now, if $v$ satisfies the same conditions we get:
\begin{align*}
|v_{L}^{*}|_{{\mathcal{H}}^{1/2}(\R)}^{2} & = \int_{\R}|\xi| | \widehat{k \ast \alpha_{L}}(\xi)|^2d\xi   \\
& =  \int_{\R}\frac{|\hat{\alpha_{L}}(\xi)|^2}{4|\xi|}d\xi\\
& =   \frac{1}{2}\int_{\R}v_{L}^{*}(t)\alpha_{L}(t)dt\\
& =   \frac{1}{2}\int_{\R}v(t)\alpha_{L}(t)dt\\
& \leq |v|_{{\mathcal{H}}^{1/2}(\R)}|v_{L}^{*}|_{{\mathcal{H}}^{1/2}(\R)}. \\
\end{align*}
\qed
 
\begin{rem}\label{rem}
The solution $v_{L}^{*}$ is in fact not defined modulo constants and coincides with $\int_{-2L}^{0}K_{L}(t,s)f(s)ds$ for all $f \in H^{1/2}(\R)$ as can be seen by letting $t$ go to $0$ or to infinity. 
\end{rem}

\subsubsection{Convergence of $v^{*}_{L,T}$ towards $v^{*}_{L}$ as $T$ goes to infinity}
We have the following exact formula:
\begin{proposition}
Let $T$ be larger than $2L$. Then for all $t$ in the interval $[0,T-2L]$, we have the identity:
\begin{equation}\label{eq:taylor}
v^{*}_{L}(t)=v^{*}_{L,T}(t)+\frac{\int_{-2L}^{0}f(s)\varphi(s/2L)ds}{\int_{-1}^{0}\varphi(s)ds} 
\frac{1-(k*\varphi)(t/2L)}{2L(1+\ln(T/2L)\int_{-1}^{0}\varphi(s)ds)}
\end{equation}

\end{proposition}
\proof
For all $t$ in the interval  $[0,T-2L]$, we have since $\int_{\R}\alpha_{L}(s)ds=0$:
\begin{align}\nonumber
v^{*}_{L}(t) & =\int_{-2L}^{0}\ln(T/|t-s|)\alpha_{L}(s)ds+\frac{\int_{-2L}^{0}f(s)\varphi(s/2L)ds}{\int_{-2L}^{0}\varphi(s/2L)ds}  \nonumber \\
& = \int_{-2L}^{0}\ln^{+}(T/|t-s|)(\alpha_{L}(s)-\alpha_{L,T}(s))ds+\int_{-2L}^{0}\ln^{+}(T/|t-s|)\alpha_{L,T}(s)ds \nonumber \\
& 
+ \frac{\int_{-2L}^{0}f(s)\varphi(s/2L)ds}{\int_{-2L}^{0}\varphi(s/2L)ds}. \label{eq:align}\\
\nonumber
\end{align}

Now, we have for all $t$ in the interval $]-2L,0[$:
\begin{equation*} 
 \int_{-2L}^{0}\ln^{+}(T/|t-s|)(\alpha_{L}(s)-\alpha_{L,T}(s))ds =-\frac{\int_{-2L}^{0}f(s)\varphi(s/2L)ds}{\int_{-2L}^{0}\varphi(s/2L)ds}.
\end{equation*}
By the uniqueness part of proposition \ref{prop:exist} and the scaling relation (\ref{eq:scale}), we get that:
\begin{equation*} 
\alpha_{L}(t)-\alpha_{L,T}(t)=-\frac{\int_{-2L}^{0}f(s)\varphi(s/2L)ds}{\int_{-2L}^{0}\varphi(s/2L)ds}\frac{\varphi(s/2L)}{2L(1+\ln(T/2L)\int_{-1}^{0} \varphi(s)ds)}
\end{equation*}
Plugging this relation into (\ref{eq:align}) yields the result.
\qed

As a corollary of the above proposition, the explicit expression (\ref{eq:explicit}) and remark \ref{rem}, we easily get theorem \ref{th:pred2}. 

\section{Appendix}
In the appendix, we show that for all $c \geq 0$, the kernel $\ln^{+}(\frac{T}{|.|+c})$ is positive (in particular, one can define a gaussian process with kernel (\ref{eq:defiX})) and we show how to derive (\ref{eq:explicit}).
\begin{lemma}\label{lem:app}
The kernel $\ln^{+}(\frac{T}{|.|+c})$ is positive. More precisely, we have the following expression for the Fourier transform if $c < T$:
\begin{equation*} 
\widehat{\ln^{+}(\frac{T}{|.|+c})}(\xi)=\frac{1}{\pi |\xi|}\int_{0}^{2\pi(T-c)|\xi|}\frac{\sin(x)}{x+2\pi|\xi|c}dx \geq 0
\end{equation*}
\end{lemma}
\proof

We have for all $\xi$:
\begin{align*}
\int_{\R}e^{-2i\pi \xi x}\ln^{+}(\frac{T}{|x|+c})dx & = \int_{-(T-c)}^{T-c}e^{-2i\pi\xi x}\ln(\frac{T}{|x|+c})dx \\
& = \int_{-(T-c)}^{0}e^{-2i\pi\xi x}\ln(\frac{T}{-x+c})dx+\int_{0}^{T-c}e^{-2i\pi\xi x}\ln(\frac{T}{x+c})dx  \\
& =\frac{1}{2i\pi \xi} \int_{-(T-c)}^{0}\frac{e^{-2i\pi\xi x}}{-x+c}dx
-\frac{1}{2i\pi \xi}\int_{0}^{T-c}\frac{e^{-2i\pi\xi x}}{x+c}dx \\
& = \frac{1}{\pi \xi}\int_{0}^{T-c}\frac{\sin(2\pi\xi x)}{x+c}dx \\
&=  \frac{1}{\pi |\xi|}\int_{0}^{2\pi(T-c)|\xi|}\frac{\sin(x)}{x+2\pi|\xi|c}dx \geq 0.
\end{align*}

\qed

In particular, one can define the gaussian process $X^{\tau}$ with covariance given by formula (\ref{eq:defiX}).

We now give a proof of identity (\ref{eq:explicit}). It is a direct consequence of the following lemma.

\begin{lemma}
For all $t$ in $]0,2[$, we have:
\begin{equation}\label{eq: conv}
\int_{0}^{2}\ln |t-s| \frac{ds}{\sqrt{2s-s^2}}=-\pi \ln(2)
\end{equation}
\end{lemma}
\proof
We start by showing that the above quantity does not depend on $t$ in the interval $]0,2[$. Since the left hand side of (\ref{eq: conv}) is invariant under $t \rightarrow 2-t$, we can consider the case 
$t \in ]0,1[$. 
By performing successively the change of variables $\sigma=\arcsin(1-s)$ and $x=\tan(\sigma/2)$, we get the following identities:
\begin{align*}
\int_{0}^{2}\ln|t-s| \frac{ds}{\sqrt{2s-s^2}} & = \int_{-\pi / 2}^{\pi /2} \ln |t-1+\sin(\sigma)| d\sigma\\
&= \frac{1}{2}  \int_{-\pi}^{\pi} \ln |t-1+\sin(\sigma)| d\sigma\\
& =\int_{\R}\ln |t-1+\frac{2x}{1+x^2}| \frac{dx}{1+x^2} \\
& =\int_{\R}\ln |(1-t)(1+x^2)-2x| \frac{dx}{1+x^2}-\int_{\R}\frac{\ln(1+x^2)}{1+x^2}dx
\end{align*}
For all $\epsilon \geq 0$, we introduce the following functions:
\begin{equation*}
\forall t \in ]0,2[, \quad F_{\epsilon}(t)=\frac{1}{2}\int_{\R}\ln(|(1-t)(1+x^2)-2x|^2+\epsilon)\frac{dx}{1+x^2}.
\end{equation*}
Note that we have to prove that $F_{0}$ is a constant function on $]0,1[$. As $\epsilon$ goes to zero,  
$F_{\epsilon}$ converges pointwise to $F_{0}$. We will prove that  $F'_{\epsilon}$ tends to 0 uniformly on compact intervals $[a,b] \subset ]0,1[$ which implies the result. If $\epsilon > 0$, we get:
\begin{equation*}
\forall t \in ]0,2[, \quad F'_{\epsilon}(t)=\int_{\R}\frac{(1-t)(1+x^2)-2x}{((1-t)(1+x^2)-2x)^2+\epsilon}dx.
\end{equation*}
Let $R$ be the rational fraction defined on $\mathbb{C}$ by:
\begin{equation*}
R(z)=\frac{(1-t)(1+z^2)-2z}{((1-t)(1+z^2)-2z)^2+\epsilon}
\end{equation*}
We get:
\begin{equation*}
R(z)=\frac{1}{(1-t)}\frac{(z-z_{+,+}(t,0))(z-z_{-,-}(t,0))}{(z-z_{+,+}(t,\epsilon))(z-z_{+,-}(t,\epsilon))(z-z_{-,+}(t,\epsilon))(z-z_{-,-}(t,\epsilon))},
\end{equation*}
where:
\begin{equation*}
z_{\pm,\pm}(t,\epsilon)=\frac{1}{1-t} \pm \sqrt{\frac{1-(1-t)^2}{(1-t)^2} \pm \frac{i\sqrt{\epsilon}}{1-t}}.
\end{equation*}
By the formula of residues, we get that for all $t$ in $]0,1[$:
\begin{align*}
F'_{\epsilon}(t) & =\frac{2\pi i}{1-t} \frac{(z_{+,+}(t,\epsilon)-z_{+,+}(t,0))(z_{+,+}(t,\epsilon)-z_{-,-}(t,0))}{(z_{+,+}(t,\epsilon)-z_{+,-}(t,\epsilon))(z_{+,+}(t,\epsilon)-z_{-,+}(t,\epsilon))(z_{+,+}(t,\epsilon)-z_{-,-}(t,\epsilon))} \\
& +\frac{2\pi i}{1-t}\frac{(z_{-,-}(t,\epsilon)-z_{+,+}(t,0))(z_{-,-}(t,\epsilon)-z_{-,-}(t,0))}{(z_{-,-}(t,\epsilon)-z_{+,+}(t,\epsilon))(z_{-,-}(t,\epsilon)-z_{+,-}(t,\epsilon))(z_{-,-}(t,\epsilon)-z_{-,+}(t,\epsilon))}).
\end{align*} 
Using the above formula, one easily derives that $F'_{\epsilon}$ goes uniformly to $0$ on $[a,b]$ as $\epsilon$ goes to $0$. 
 
 Since $F_{0}$ is a constant function on $]0,2[$, we only need to compute $F_{0}(1)$ to prove (\ref{eq: conv}).  The computation of  $F_{0}(1)$ is standard.

\qed

\textbf{Acknowledgements}: The authors would like to thank Capital Fund Management (CFM) for providing them data on the SP500 Index and the currency EUR/AUD. V. Vargas would like to thank J.P. Bouchaud and the Nimbus team at CFM for useful discussions.

\newlength{\figlen}
\setlength{\figlen}{13cm}

\begin{figure}[htp]
   \includegraphics[width=\figlen]{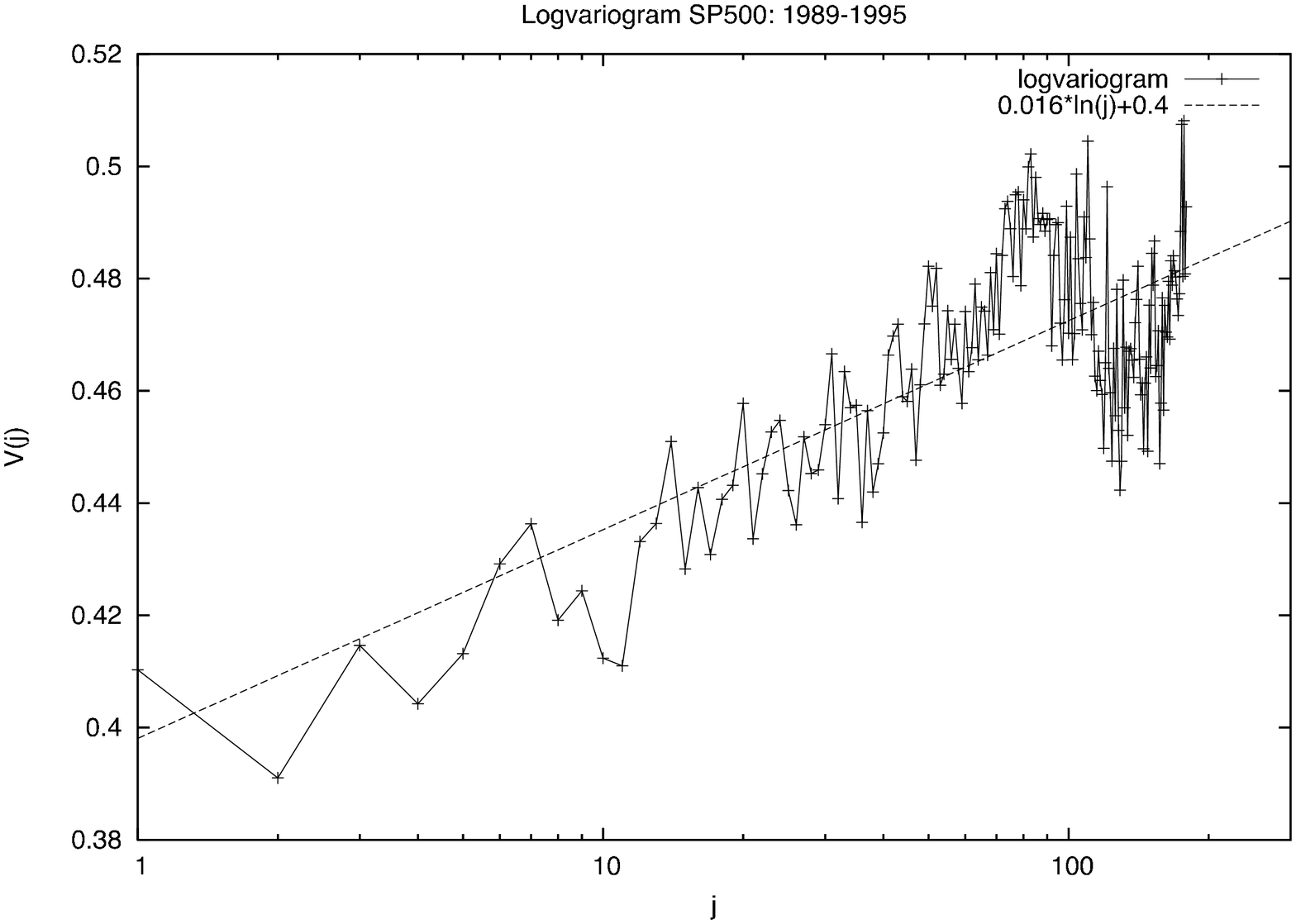}
   \includegraphics[width=\figlen]{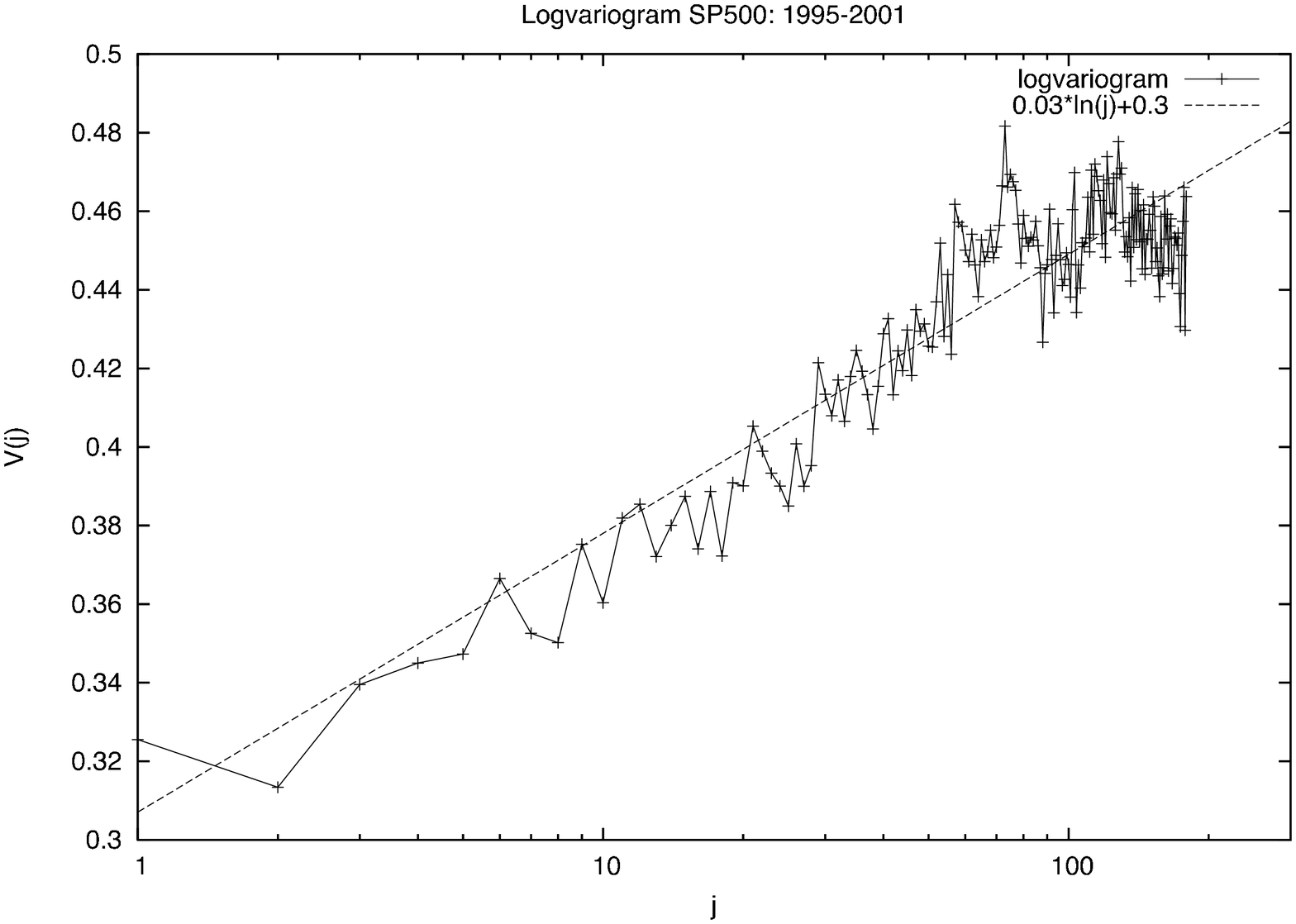}
   \caption{\textbf{First plot}: empirical logvariogram of the SP500 index on the period 1989-1995. \textbf{Second plot}: empirical logvariogram of the SP500 index on the period 1995-2001.}
   \label{fig:first}
\end{figure}

\begin{figure}[htp]
  \includegraphics[width=\figlen]{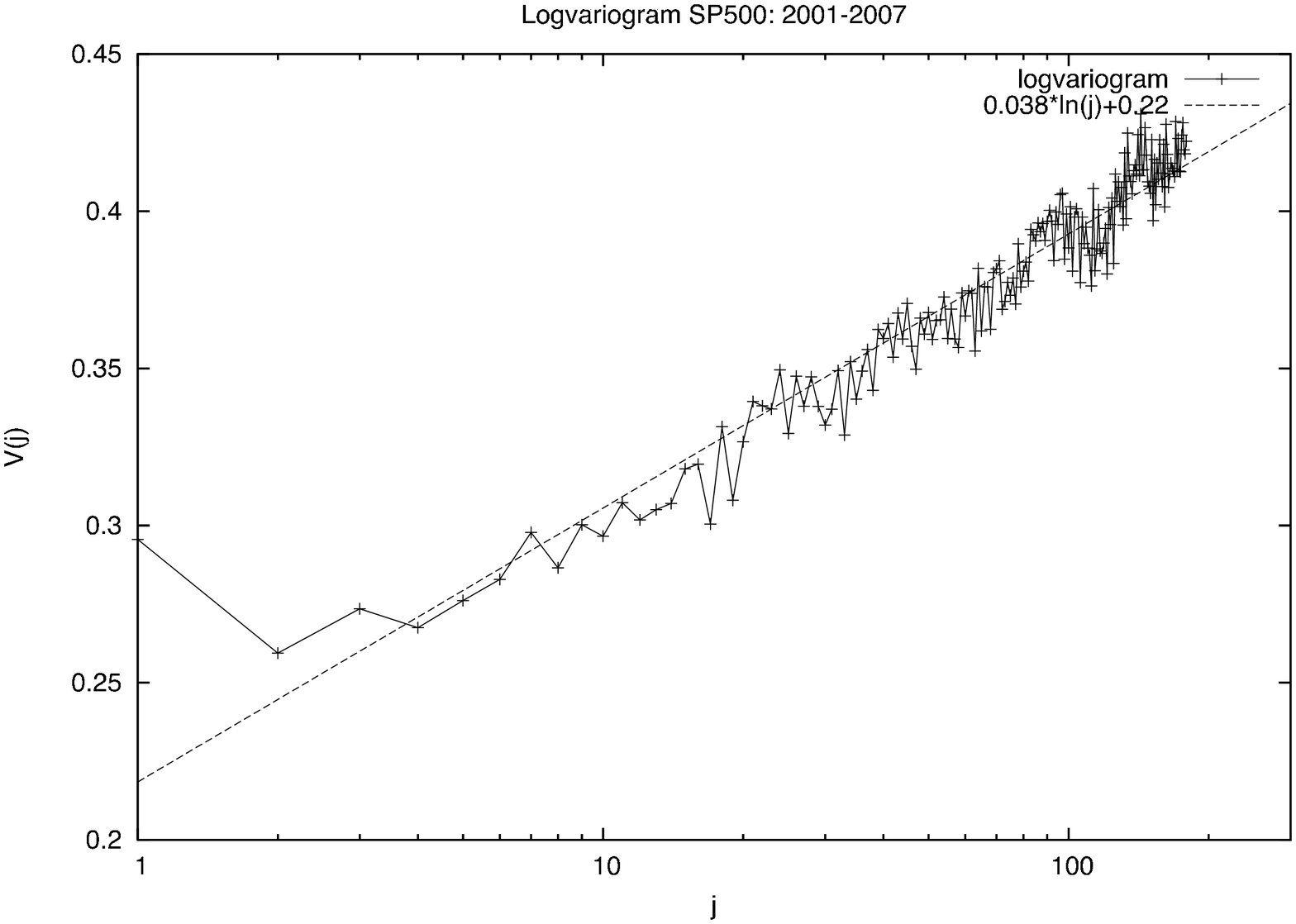}
  \includegraphics[width=\figlen]{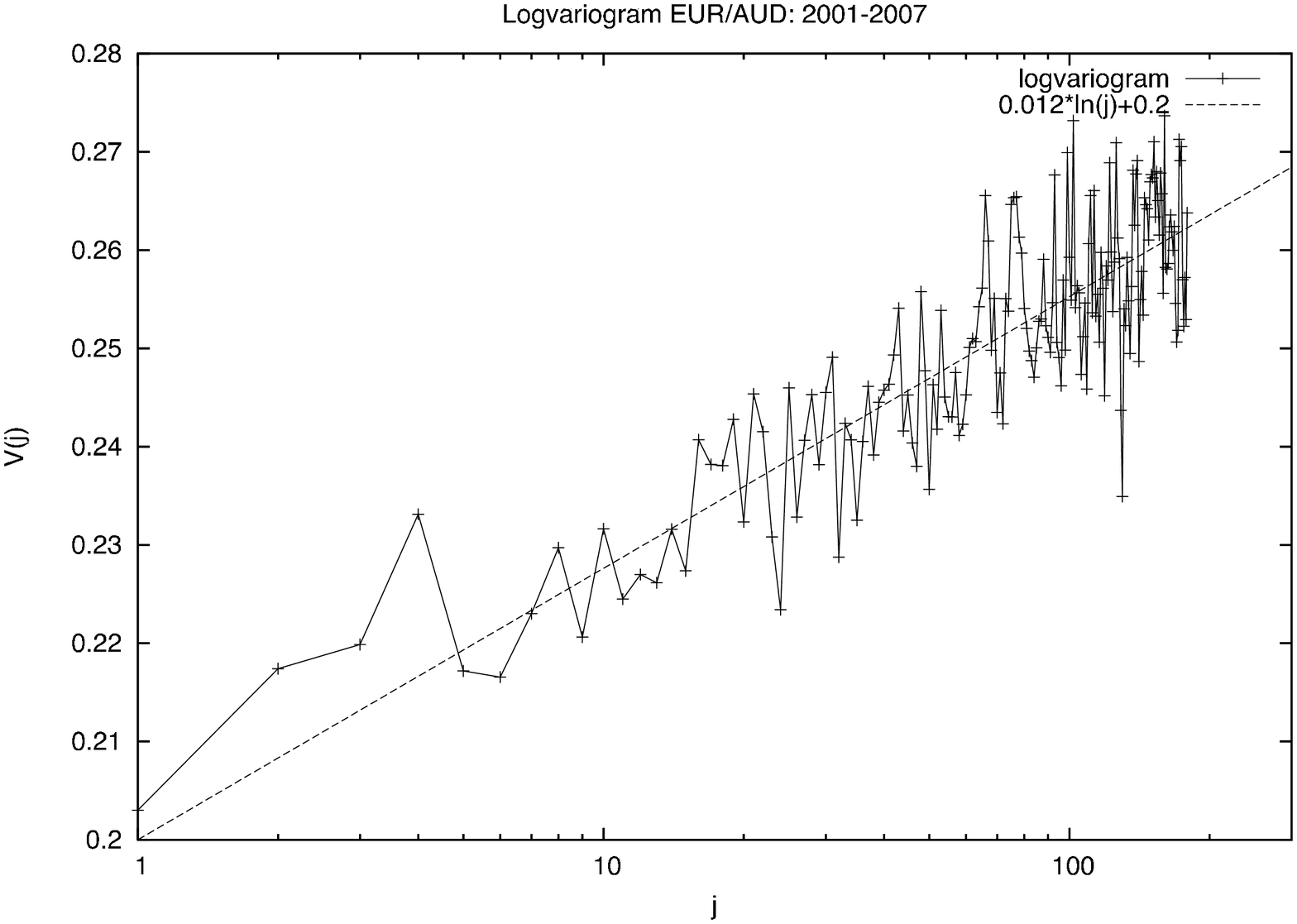}
   \caption{\textbf{First plot}: empirical logvariogram of the SP500 index on the period 2001-2007. \textbf{Second plot}:
   empirical logvariogram of the currency EUR/AUD on the period 2001-2007 in the London time zone 
   (Bloomberg series EUR/AUD CMPL)
    }
   \label{fig:second}
\end{figure}


\end{document}